\documentclass[aps,prd,twocolumn,superscriptaddress,floatfix,nofootinbib,showpacs,amsmath,amssymb,altaffilletter]{revtex4-2}
\pdfoutput=1

\UseRawInputEncoding

\usepackage{amsmath}
\usepackage{amsfonts}
\usepackage{amssymb}
\usepackage{graphicx}
\usepackage{bm}
\usepackage{subfigure}
\usepackage{url}
\usepackage[hyperindex]{hyperref}
\usepackage{color}
\usepackage[ddmmyy,24hr]{datetime}
\usepackage{bigdelim}
\usepackage{booktabs}
\usepackage{dcolumn}
\usepackage{multirow}
\usepackage{subfigure}
\usepackage{physics}
\usepackage{cancel}
\usepackage{stackrel}
\usepackage{paralist}
\usepackage{xspace}
\usepackage{slashed}
\usepackage{cancel}
\usepackage{todonotes}
\usepackage{enumerate}
\usepackage{float}
\usepackage{fullpage}
\usepackage{ulem}
\usepackage{wasysym}
\usepackage{comment}
\usepackage{bbold}
\usepackage{orcidlink}

\usepackage{feynmp}
\newcommand{\nua}[1]{\ensuremath{\rlap{\kern-2.5pt\ensuremath{\overset{\scriptscriptstyle(-)}{\phantom{\nu}}}}{\ensuremath{{\nu}_{#1}}}}}

\usepackage{enumitem}

\begin{document}

\title{Reassessing the gallium anomaly using self-consistent electron wave functions}

\author{M. Cadeddu \orcidlink{0000-0002-3974-1995}}
\email{matteo.cadeddu@ca.infn.it}
\affiliation{Istituto Nazionale di Fisica Nucleare (INFN), Sezione di Cagliari,
	Complesso Universitario di Monserrato - Strada Provinciale per Sestu Km 0.700,
	09042 Monserrato (Cagliari), Italy}

\author{N. Cargioli \orcidlink{0000-0002-6515-5850}}
\email{nicola.cargioli@ca.infn.it}
\affiliation{Istituto Nazionale di Fisica Nucleare (INFN), Sezione di Cagliari,
	Complesso Universitario di Monserrato - Strada Provinciale per Sestu Km 0.700,
	09042 Monserrato (Cagliari), Italy}

\author{G. Carotenuto }
\email{giovanni.carotenuto@terna.it}
\affiliation{Dipartimento di Fisica, Universit\`{a} degli Studi di Cagliari,
	Complesso Universitario di Monserrato - Strada Provinciale per Sestu Km 0.700,
	09042 Monserrato (Cagliari), Italy}

\author{F. Dordei \orcidlink{0000-0002-2571-5067}}
\email{francesca.dordei@cern.ch}
\affiliation{Istituto Nazionale di Fisica Nucleare (INFN), Sezione di Cagliari,
	Complesso Universitario di Monserrato - Strada Provinciale per Sestu Km 0.700,
	09042 Monserrato (Cagliari), Italy}

\author{L. Ferro \orcidlink{0009-0002-1698-3710}}
\email{luca.ferro@ca.infn.it}
\affiliation{Dipartimento di Fisica, Universit\`{a} degli Studi di Cagliari,
	Complesso Universitario di Monserrato - Strada Provinciale per Sestu Km 0.700,
	09042 Monserrato (Cagliari), Italy}
\affiliation{Istituto Nazionale di Fisica Nucleare (INFN), Sezione di Cagliari,
	Complesso Universitario di Monserrato - Strada Provinciale per Sestu Km 0.700,
	09042 Monserrato (Cagliari), Italy}

\author{C. Giunti \orcidlink{0000-0003-2281-4788}}
\email{carlo.giunti@to.infn.it}
\affiliation{Istituto Nazionale di Fisica Nucleare (INFN), Sezione di Torino, Via P. Giuria 1, I--10125 Torino, Italy}


\begin{abstract}
The gallium anomaly, a persistent discrepancy exceeding $4\sigma$ in the $^{71}$Ga neutrino capture rates from $^{51}$Cr and $^{37}$Ar radioactive sources by the GALLEX, SAGE, and recently BEST experiments, has challenged particle physics and nuclear theory for over three decades. We present a new calculation of the neutrino capture cross section, abandoning the conventional leading-order approximation for electronic wave functions by numerically solving the Dirac-Coulomb equation for both bound and continuum electron states. 
Finally, we reevaluate the gallium anomaly, updating its global significance and presenting the most up-to-date status of its interpretation in terms of sterile neutrinos.
\end{abstract}

\maketitle

\section{Introduction}
The study of neutrino interactions with matter has provided profound insights into both the properties of neutrinos and the dynamics of weak interactions. In this field, a particularly intriguing and long-standing puzzle is the so-called \textit{gallium anomaly}~\cite{Elliott:2023cvh, Giunti:2010zu}, observed in experiments involving gallium-71 ($^{71}{\textrm{Ga}}$) as a neutrino target. This anomaly, identified at the GALLEX~\cite{ANSELMANN1992376,1999127,Kaether:2010ag,GNO:2005bds} and SAGE~\cite{SAGE:2009eeu,doi:10.1142/9789811204296_0002} experiments, which reported a measured neutrino-induced inverse beta decay (IBD) rate from $^{51}$Cr and $^{37}$Ar calibration sources systematically lower than predictions, has raised questions about the robustness of the theoretical cross section calculations. Moreover, when considered alongside other anomalies~\cite{Mention:2011rk,MicroBooNE:2022sdp,Giunti:2022btk,Berryman:2019hme}, it has strengthened the case for potential implications beyond the standard model~\cite{Giunti:2006bj,Giunti:2019aiy,Gariazzo:2015rra,Gonzalez-Garcia:2015qrr,Diaz:2019fwt, Boser:2019rta,Dasgupta:2021ies,Giunti:2023kyo,Abazajian:2012ys,Acero:2022wqg,Farzan:2023fqa}.

The anomaly arises in the charge current \mbox{$\nu_e + ^{71}{\textrm{Ga}} \to \, ^{71}{\textrm{Ge}} + e^-$} reaction, a process historically used to detect neutrinos from solar and artificial sources. While the theoretical predictions for the cross section have provided critical benchmarks for interpreting experimental results, the persistent deficit in the observed rates, now exceeding $4\sigma$ thanks to the recent short-baseline BEST experiment~\cite{Barinov:2021asz,PhysRevC.105.065502}, demands a closer examination of the underlying assumptions and parameters in the theoretical models. An early theoretical work by Bahcall~\cite{Bahcall:1997eg} established the foundation for these predictions. However, the persistence of this anomaly later prompted other theoretical investigations~\cite{Elliott:2023xkb,Haxton:2025hye,Kostensalo:2019vmv,Brdar:2023cms,Giunti:2022xat,Huber:2022osv} that employed more recent experimental inputs, the inclusion of radiative terms, additional corrections, and an updated evaluation of excited-state contributions, among others. 

Our analysis builds on and updates these previous works. In this paper, we undertake a comprehensive revision of the theoretical cross section for neutrino interactions with $^{71}{\textrm{Ga}}$.
The widely employed theoretical approach relies on the \textit{detailed balance} (db) principle~\cite{Williams,Blatt:1952ije, Bahcall:book,Alvarez:1949zz} between neutrino-induced IBD and electron-capture (EC) rates, factorizing nuclear and leptonic matrix elements, and utilizing leading-order (LO) approximations for the 
lepton
wave functions~\cite{Bambynek, Behrens}. 
Here, we abandon the conventional LO approximation for electronic wave functions by numerically solving the Dirac-Coulomb equation for both bound and continuum electron states, using the same underlying assumptions and software.  The latter is used to obtain an accurate determination of the Fermi function~\cite{Fermi:1933jpa}, a correction for the Coulomb distortion of the outgoing electron in the field of the associated nucleus.
While doing so, we incorporate recent developments in the evaluation of the electron-capture lifetime of $^{71}{\textrm{Ge}}$~\cite{PhysRevC.31.666,Collar:2023yew,PhysRevC.109.055501,newGelifetime} and the first determination of the nuclear charge radius of $^{71}{\textrm{Ge}}$~\cite{ExpNuclearRadius}, the use of a more accurate \mbox{\textit{Q}-value}~\cite{Qvalue} as well as updated electron-capture probability ratios~\cite{Collar:2023yew, PhysRevLett.125.141301,Bambynek}. 

These updates aim to provide a more accurate and self-consistent theoretical framework for interpreting gallium-based neutrino experiments. Moreover, we update the status of the allowed parameter space for the interpretation of such an anomaly in terms of sterile neutrinos~\cite{Elliott:2023cvh,Giunti:2019aiy,Giunti:2006bj,Boser:2019rta} whose presence would result in a reduction of the observed IBD event rate.

\section{Determination of the neutrino capture cross section}
The ground-state neutrino-induced IBD cross section of the process $\nu_e + ^{71}\textrm{Ga} \to \, ^{71}{\textrm{Ge}} + e^-$  can be written as\footnote{Natural units with $\hbar = c = 1$ are used throughout.} \cite{BahcallRevModPhys, Bahcall:1997eg,Bahcall:book,Bahcall1964}
\begin{equation}
\sigma_\text{gs}=\frac{G_F^2\,|V_{ud}|^{2}\,g^2_{A}\,\Omega_e \Omega_\nu}{\pi (2J_\text{Ga}+1)}\sum_{j}{ {p^j_e E^j_e}\,
\left|\mathcal{H}_j^{\rm IBD}\right|^2 \,\mathcal{B}(E^j_e)},
\label{eq:sigma_IBD}
\end{equation}
where $G_F$ is the Fermi constant, $V_{ud}$ is the Cabibbo-Kobayashi-Maskawa matrix element~\cite{ParticleDataGroup:2024cfk}, \mbox{$g_A=1.2764$} is the axial-vector coupling constant~\cite{PERKEO3}, $\Omega_e$ and $\Omega_\nu$ are normalization volumes and $J_\text{Ga}=3/2$ is the nuclear spin of gallium. The momentum of the electron is indicated by $p_e$, while $E_e$ is the energy of the emitted electron, which is related to the $Q$-value of the process, $Q_\text{EC}$, and to the energy of the absorbed neutrinos by $E_e = E_\nu - Q_\text{EC} + m_e - 0.09\,\text{keV}$.\footnote{Following Ref.~\cite{Bahcall:1997eg}, we consider a small correction of \mbox{0.09 keV} representing the energy lost during the electronic rearrangement.} Finally, the sum runs over the different possible electron energies, $E_e^j$, weighted by the corresponding branching ratio, $\mathcal{B}(E^j_e)$, of the neutrino source employed in gallium experiments. The term $\mathcal{H}_j^{\rm IBD}$ represents the matrix element of the transition and incorporates the involved lepton, $\psi$, and nuclear, $\Psi$, wave functions,
and the Gamow-Teller Hamiltonian, $\hat{H}_{\text{GT}}$. \\
In the literature, it is common practice to assume that the neutrino can be considered to be a plane wave normalized in such a way to cancel $\Omega_\nu$ in Eq.~(\ref{eq:sigma_IBD}), and that the electron wave functions vary slowly within the nuclear volume. Therefore, the latter can be factored out by considering its value at a specific distance $r_0$, usually chosen to be $r_0=0$ or $r_0=R_{\rm box}$, where $R_{\rm box}$ represents the equivalent box density radius, namely, $R_{\rm box}=\sqrt{5/3}\,R_{\rm ch}$, with $R_{\rm ch}$ being the nuclear charge radius. In this way, one can isolate the IBD nuclear matrix element, $\mathcal{M}^{\rm IBD}_{\rm nuc}$,\footnote{In the literature, the nuclear matrix element is often referred to in terms of the Gamow-Teller strength \mbox{$B_{\rm GT}=|\mathcal{M}_{\rm nuc}|^2/(2 J+1)$}, with $J$ being the total angular momentum of the initial nuclear state}.
\begin{equation}
    \mathcal{H}^{\rm IBD}_j\simeq\psi^{j\,*}_{e}(r_0)\,\mathcal{M^{\rm IBD}_{\rm nuc}}\, ,
    \label{eq:Hmibd}
\end{equation}
where
\begin{equation}
    \mathcal{M^{\rm IBD}_{\rm nuc}}=\int
\Psi^*_{{}^{71}\text{Ge}}(\mathbf{r})\,\hat{H}_{\text{GT}}\,
\Psi_{{}^{71}\text{Ga}}(\mathbf{r})\,\text{d}\mathbf{r}\, .
\label{eq:mibd}
\end{equation}

This permits us to use the detailed balance  principle, according to which the nuclear matrix element in Eq.~(\ref{eq:mibd}) can be extracted from the inverse process, \mbox{$
e^-_{\text{bound}} + {}^{71}\text{Ge} \rightarrow {}^{71}\text{Ga} + \nu_e$}, thus from the measurement of the electron-capture lifetime of $^{71}{\textrm{Ge}}$, referred to as $t_{1/2}$.
The electron-capture rate is defined as~\cite{Bambynek, Behrens,BahcallEC1962,Bahcall1964,BryskRose}
\begin{equation}
\lambda_\text{EC}\!=\! \frac{\ln 2}{t_{1/2}}\!=\!\frac{G_F^2\,|V_{ud}|^{2}\,g^2_{A}}{2\pi (2J_\text{Ge}\!+\!1)}
\sum_{x}\! n_x (E^x_{\nu})^2\, |\mathcal{H}^{\rm EC}_x|^2\,,\label{eq:lambdaEC_iniz}
\end{equation}

with $J_\text{Ge}=1/2$ being the nuclear spin of the ground state of $^{71}{\textrm{Ge}}$, $n_x$ being the relative occupation number of electrons in the atomic $x$ shell, and $\mathcal{H}^{\rm EC}_x$ being the EC matrix element. In both the IBD cross section in Eq.~(\ref{eq:sigma_IBD}) and the EC rate in Eq.~(\ref{eq:lambdaEC_iniz}) we account for radiative corrections and the weak magnetism contribution following Ref.~\cite{Elliott:2023xkb}. Factorizing out the leptonic wave functions as in the IBD case, the electron-capture matrix element becomes
\begin{align}
\mathcal{H}^\text{EC}_x\simeq\psi^{\rm b}_{e,x}(r_0)\,\mathcal{M^{\rm EC}_{\rm nuc}}\, ,
\label{eq:lambdaEC}
\end{align}
where we introduced the superscript \textit{b} to the electron wave function to emphasize the electrons being bound in the EC process, contrary to the IBD case. 

In this way, the EC nuclear matrix element can be retrieved as
\begin{equation}
\frac{|\mathcal{M^{\rm EC}_{\rm nuc}}|^2}{2J_\text{Ge}+1} = \dfrac{2\pi^3\,\ln 2}{G_F^2\,|V_{ud}|^{2}\,g^2_{A}\, f_\text{EC}\,t_{1/2}}\, ,
\label{eq:matrice EC}
\end{equation}
with $f_\text{EC}$ being the phase space factor for an allowed electron capture, which, given that EC happens mainly with the $1s$ shell electrons, can be expanded as \cite{BahcallRevModPhys, Bahcall:book} 
\begin{equation}
    f_\text{EC}
    \!\simeq\! 2\pi^2|\psi^{\rm b}_{e,1s}(r_0) |^2 (E_{\nu}^{1s})^2 (1+\epsilon_\text{o}^{1s})\! \left[ 1 \!+\! {P_\text{L}+P_\text{M} \over P_\text{K}}\!\right],
\label{eq:fEC}
\end{equation}
with $\epsilon_\text{o}^{1s}$ accounting for exchange and overlap effects, 
and $P_\text{L,M,K}$ being the experimentally measured electron-capture probabilities for the L, M and K shells~\cite{Collar:2023yew, PhysRevLett.125.141301,Bambynek}.
Here, $|\psi^{\rm b}_{e,1s}(r_0)|^2$ is the $1s$ electron density at the nucleus obtained from 
\begin{equation}
|\psi^{\rm b}_{e,1s}(r)|^2=\dfrac{|g_{1s}(r)|^2+|f_{1s}(r)|^2}{4\pi}\, ,\label{eq:psi_bound}
\end{equation}
where $g_{1s}(r)$ and $f_{1s}(r)$ represent the exact solution of the Dirac-Hartree-Fock-Slater (DHFS) equations~\cite{radial} for a bound electron in the $1s$ state.

Following the detailed balance principle, $|\mathcal{M}^{\rm IBD}_{\rm nuc}|^2\stackrel{\rm db}{=}|\mathcal{M}^{\rm EC}_{\rm nuc}|^2$,
the ground-state neutrino-induced IBD cross section reduces to~\cite{Elliott:2023xkb}
\begin{equation}
\sigma^{\rm db}_{\mathrm{gs}} \!=\! \!\frac{2\pi^2\! \ln2}{f_\text{EC}\,t_{1/2}}\!\left(\frac{2J_\text{Ge}\!+\!1}{2J_\text{Ga}\!+\!1} \right)\!
\!\sum_{j}p^j_e E^j_e
\mathcal{F}(E_e,Z,r_0)\mathcal{B}(E^j_e),
\label{eq:sigmaIBDInt}
\end{equation}
where, to match the usual conventions, we have normalized the nondistorted electron wave function, $|\psi_e(r_0)|_\text{free}^2$, to cancel $\Omega_e$ in Eq.~(\ref{eq:sigma_IBD}), and introduced the Fermi function, \mbox{$\mathcal{F}(E_e,Z)=|\psi_e(r_0)|^2/|\psi_e(r_0)|^2_\text{free}$}, which corrects for the distortion of the electron wave function due to the Coulomb potential of the daughter nucleus, with $Z$ being its atomic number. 
Here, we use a generalized Fermi function
given by~\cite{Hayen:2017pwg} 
\begin{equation}
\mathcal{F}(E_e, Z,r) = \frac{|f_1(r)|^2 + |g_{-1}(r)|^2}{2p_e^2},
\label{eq:fermi}
\end{equation}
where the electron wave functions $f_\kappa(r)$ and $ g_\kappa(r)$ are the exact radial solutions of the DHFS equations for the outgoing electron of the IBD process for the eigenvalue $\kappa$, which assumes values $\kappa = \pm 1$ for allowed decays. 

In this work, we developed a software based on the \texttt{RADIAL} package~\cite{radial} that allows us to numerically evaluate the Fermi function in Eq.~(\ref{eq:fermi}), using the exact DHFS radial electron wave functions, $\psi_{e,\kappa}(\mathbf{r})$, that are given in terms of  
the large and small components, $g_\kappa(r)$ and $f_\kappa(r)$, 
obtained by solving
\begin{eqnarray}
		\label{eq:eqDirac}
		\left(\frac{d}{dr}+\frac{\kappa+1}{r}\right)g_\kappa-(E_e-V(r)+m_e)f_\kappa=0,\,\,\,\,\,\,&&\nonumber\\
		\left(\frac{d}{dr}-\frac{\kappa-1}{r}\right)f_\kappa+(E_e-V(r)-m_e)g_\kappa=0,\,\,\,\,\,\,&&
	\end{eqnarray}
where $V(r)$ is the atomic potential of the germanium atom.
\begin{figure}[t]
    \centering
\includegraphics[width=1\linewidth]{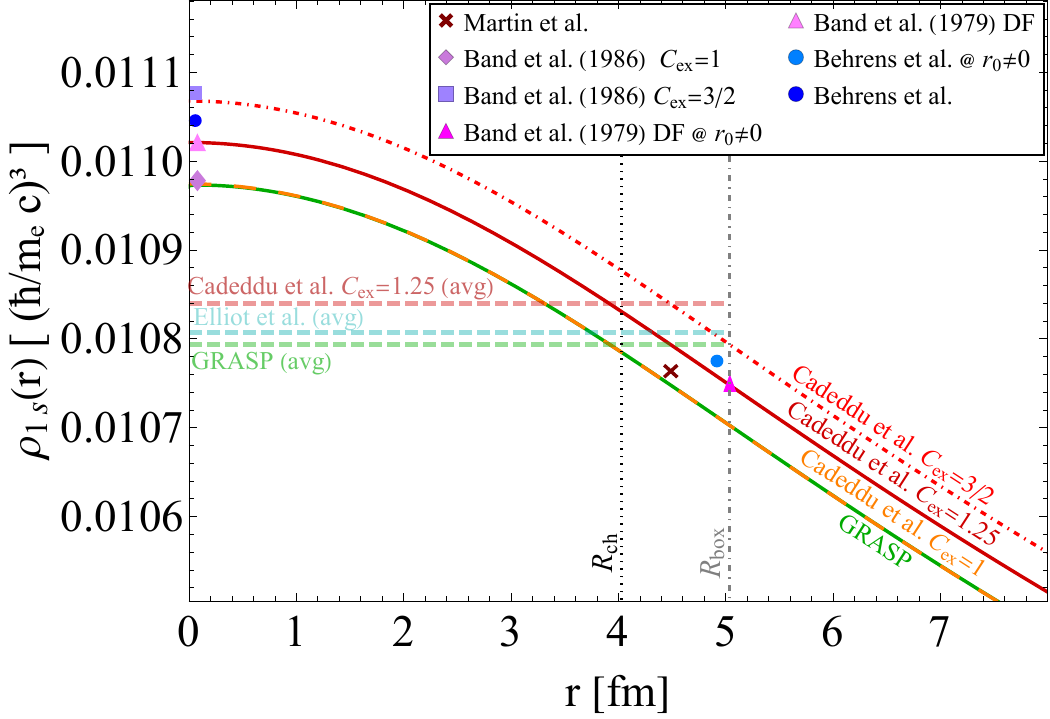}
    \caption{Electron density at the nucleus as a function of the radial distance from the origin. We compare our calculations based on \texttt{RADIAL}~\cite{radial} (labeled ``Cadeddu \textit{et al.}'') with that obtained using \texttt{GRASP}~\cite{Grasp2018} and with those available in the literature~\cite{Martin,BAND1986,BAND1979,Behrens}. The horizontal dashed lines indicate the values obtained after applying the averaging procedure. The vertical dotted and dash-dotted lines represent $R_{\mathrm{ch}}$ and \mbox{$R_{\mathrm{box}}=\sqrt{5/3}\, R_{\rm ch}$}, respectively.}
    \label{fig:eledensityComparison}
\end{figure}
Similar functions, $\psi^{\rm b}_{e,n\kappa}(\mathbf{r})$, $g_{n\kappa}(r)$ and $f_{n\kappa}(r)$, can be defined for the case of bound electron states, with the replacement of the electron energy $E_e$ with $E_e^{n\kappa}=m_e-|E_{\rm bind}^{n\kappa}|$, where $E_{\rm bind}^{n\kappa}$ is the binding energy and introducing a subscript $n$ indicating the principal quantum number.
In our work, we consider the DHFS potential, which is given by 
\begin{equation}
V_\text{DHFS}(r) = V_\text{nuc}(r) + V_\text{el}(r) + V_\text{ex}(r)\,,
    \label{eq:total_potential}
\end{equation}
to which we apply the so-called Latter tail correction~\cite{LatterPhysRev.99.510,radial}, in order to ensure the correct long-range behavior.
Here, $V_\text{nuc}(r)$ represents the nuclear potential, which has been determined by considering a two-parameter Fermi (2pF) distribution~\cite{PhysRevLett.120.072501} with the first measurement of the charge radius of $^{71}{\textrm{Ge}}$ as an input parameter, equal to  \mbox{$R_{\rm ch} = 4.032\pm0.002$~fm~\cite{ExpNuclearRadius}}, 
and skin thickness \mbox{$t=2.3\, \mathrm{fm}$}. The electronic potential $V_\text{el}(r)$ accounts for the interaction energy of an electron with the entire electronic cloud at a distance $r$, and the exchange potential $V_\text{ex}(r)$ is defined as
\begin{equation}
    V_\text{ex}(r)=-C_\text{ex}e^2\bigg(\frac{3}{\pi}\bigg)^{1 / 3}\,[\rho(r)]^{1/3}\,,
    \label{eq:exchange_potential}
\end{equation}
where $\rho(r)$, which enters both $V_{\rm el}(r)$ and $V_{\rm ex}(r)$, is determined through a self-consistent solution of the DHFS equations iteratively~\cite{radial}.
The constant $C_\text{ex}$, as discussed in Refs.~\cite{Cex,PhysRev.171.1}, varies according to the chosen potential model. For example, \mbox{$C_\text{ex} = 3/2$} corresponds to the Slater approximation~\cite{SlaterPhysRev.81.385,radial} while $C_\text{ex} =1$ to the Kohn-Sham one~\cite{KhonPhysRev.140.A1133,radial}.
\begin{table*}[t]
\resizebox{0.96\textwidth}{!}
{\renewcommand{\arraystretch}{1.25} 
\begin{tabular}{l|c|c|c}
    Reference  & $\rho_{\rm ch}(r)$ & $\rho_{\rm ch}(r)$ input par.$\,[\text{fm}]$ & $4\pi\,|\psi^{\rm b}_{e,1s}(r_0)|^2$ \\
\hline 
    Band \textit{et al.} (1979)~\cite{BAND1979} ($r_0=0$) DF\footnote{Here, DF refers to the use of a different resolution method, namely Dirac-Fock instead of the DHFS one, for which the exchange effect is evaluated without any approximation. The same result has also been reported in Ref.~\cite{BAND1986}.} & Box [$\text{A=74}$]& $1.2\,\text{A}^{1/3}$ & 0.0692  \\
    Band \textit{et al.} (1986)~\cite{BAND1986} ($r_0=0$) for $C_{\rm ex}=\frac{3}{2}$
    & Box [$\text{A=74}$]& $1.201\,\text{A}^{1/3} $ & 0.0696 \\
    Band \textit{et al.} (1986)~\cite{BAND1986} ($r_0=0$) for $C_{\rm ex}=1$& Box [$\text{A=74}$]& $1.201\,\text{A}^{1/3} $ & 0.0690 \\
    Behrens \textit{et al.}~\cite{Behrens} ($r_0=0$) & Box [$\text{A=69}$]& $1.2\,\text{A}^{1/3}$ & 0.0694 \\
    Bambynek \textit{et al.}~\cite{Bambynek} ($r_0=0$) & 2pF [A=71] & $c=4.43,\, t=2.4$ & 0.0692 \\
    \hline
   Band \textit{et al.} (1979)~\cite{BAND1979} ($r_0=5.04\,\mathrm{fm}$) DF & Box [$\text{A=74}$]& $1.2\,\text{A}^{1/3}$ & 0.0675 \\
    Behrens \textit{et al.}~\cite{Behrens} ($r_0=4.92\,\mathrm{fm}$) & Box [$\text{A=69}$]& $1.2\,\text{A}^{1/3}$ & 0.0677 \\
    Martin \textit{et al.}~\cite{Martin} ($r_0=4.49\,\mathrm{fm}$)\footnote{Ref.~\cite{Martin} quotes $|g_{1s}|^2$ and does not report the value of A. Thus, we consider  A=74 which corresponds to the most abundant stable isotope of germanium.} & 2pF\,[$\text{A=74}$] & \parbox{5cm}{ \centering $c=4.49, \,t=2.5$} & 0.0676 \\
    Elliott \textit{et al.}~\cite{Elliott:2023xkb} (avg.)\footnote{Ref.~\cite{Elliott:2023xkb} uses the same wave functions reported by Bahcall in Ref.~\cite{Bahcall:1997eg}, 
     after applying, in their calculations, corrections to the neutrino energy.} & 2pF [A=71] & -- & 0.0679 \\\hline \hline
    \texttt{GRASP}~\cite{Grasp2018} ($r_0=0$) & 2pF [A=71] & $c=4.559,\, t=2.3$ & 0.0689 \\
    \texttt{GRASP}~\cite{Grasp2018} ($r_0=5.20\,\text{fm}$) & 2pF [A=71] & $c=4.559,\, t=2.3$ & 0.0671  \\
    \texttt{GRASP}~\cite{Grasp2018} (avg.) & 2pF [A=71] & $c=4.559,\, t=2.3$ & 0.0678
    \\\hline \hline
    Cadeddu \textit{et al.} ($r_0=0$) for $C_{\rm ex}=\frac{3}{2}$ & 2pF [A=71] & $c=4.559,\, t=2.3$ & 0.0695 \\
    Cadeddu \textit{et al.} ($r_0=0$) for $C_{\rm ex}=1$ & 2pF [A=71] & $c=4.559,\, t=2.3$ & 0.0689 \\
    Cadeddu \textit{et al.} ($r_0=0$) for $C_{\rm ex}=1.25$ & 2pF [A=71] & $c=4.559,\, t=2.3$ & 0.0692 \\
    Cadeddu \textit{et al.} ($r_0=5.20\,\text{fm}$) for $C_{\rm ex}=\frac{3}{2}$& 2pF [A=71] & $c=4.559,\, t=2.3$ & 0.0677  \\
    Cadeddu \textit{et al.} ($r_0=5.20\,\text{fm}$) for $C_{\rm ex}=1$& 2pF [A=71] & $c=4.559,\, t=2.3$ & 0.0671  \\
    Cadeddu \textit{et al.} ($r_0=5.20\,\text{fm}$) for $C_{\rm ex}=1$.25 & 2pF [A=71] & $c=4.559,\, t=2.3$ & 0.0674  \\
    \textbf{Cadeddu \textit{et al.} (avg.) for $\mathbf{C_{\rm ex}=1.25}$} & \textbf{2pF [A=71]} & $\mathbf{c=4.559,\, t=2.3}$ & \textbf{0.0681}\\
\end{tabular}
}
\caption{
Comparison of different electron densities for the $1s$ bound state of $^{71}$Ge at a specific radius $r_0$, $4\pi\,|\psi^{\rm b}_{e,1s}(r_0)|^2$, according to various calculations available in the literature, which consider nuclear charge densities, $\rho_{\rm ch}(r)$. ``A" refers to the considered germanium mass number, which typically enters in the definition of the charge radius.  The table presents results from various references, specifying $\rho_{\rm ch}(r)$ (box or 2pF distribution), the associated input parameters, and the corresponding values of $4\pi\,|\psi^{\rm b}_{e,1s}(r_0)|^2$, in units of $(\hbar/m_e c)^3$. We also report the electron density, as obtained using a different Dirac equation solver, \texttt{GRASP}~\cite{Grasp2018}. In the last seven rows, we show the values obtained in this work for $r_0=0$, $r_0=\sqrt{5/3}R_{\rm ch}$ and with the averaging procedure (avg) considering  different values of $C_{\rm ex}$ entering Eq.~(\ref{eq:exchange_potential}).}
\label{tab:wavefunction_comparison}
\end{table*}
DHFS calculations neglect additional effects such as the Breit interaction, vacuum polarization, and self-energy corrections, which, however, are expected to be subdominant for the problem under investigation. In Fig.~\ref{fig:eledensityComparison}, we compare our results based on \texttt{RADIAL} to those obtained exploiting the \texttt{GRASP}~\cite{Grasp2018} software as well as to other results available in the literature in terms of $\rho_{\rm 1s}(r)=2 |\psi^{\rm b}_{e,1s}(r)|^2$, where the factor 2 represents the electron multiplicity. \texttt{GRASP} solves the Dirac equation without any approximation for the exchange effects, including also the aforementioned contributions. Interestingly, the results obtained using \texttt{RADIAL} for $C_{\rm ex}=1$ are in perfect agreement with those of \texttt{GRASP}. Moreover, we observe that a variation of $C_{\rm ex}$ produces a difference in the electron density at the nucleus smaller than 1\%, in agreement with the findings of Ref.~\cite{BAND1986}.
Given that $C_{\rm ex}$ is expected to assume a value between \mbox{$C_\text{ex} = 3/2$} and $C_\text{ex} =1$, we chose to fix its values to the average one, namely $C_{\rm ex}=1.25$, adding a systematic uncertainty to account for the observed spread. The density at the nucleus corresponding to this choice is also shown in Fig.~\ref{fig:eledensityComparison}, from which it can be compared to the Dirac-Fock result obtained in Refs.~\cite{BAND1979,BAND1986}, finding an excellent agreement.
The numerical values obtained for the electron density are reported in Table~\ref{tab:wavefunction_comparison} in units of $(\hbar/m_e c)^3$, along with those available in the literature.  
We checked that, when using similar assumptions, the electron densities evaluated with our code are compatible with those reported in the literature within about 0.2\%.\\

Besides the numerical evaluation of the electron density at the nucleus, we employed the same dedicated software, based on the \texttt{RADIAL} package~\cite{radial}, also to numerically evaluate the free electron wave functions for a finite-size nucleus, which enter the generalized definition of the Fermi function in Eq.~(\ref{eq:fermi}). This has to be compared to the common definition of the Fermi function, which can be found in the literature (see, e.g., Ref.~\cite{Elliott:2023xkb}), obtained by the product of various contributions, namely,
\begin{equation}
F(E_e,Z)
 \!=\!F_0(E_e,Z) L_0(E_e,Z)U(E_e,Z)S(E_e,Z),
\label{eq:Fermi function_lit}
\end{equation} 
where $F_0(E_e,Z)$ is derived from the solution of the Dirac equation for a pointlike nucleus. The latter is typically evaluated at the nuclear surface to keep it finite, and its expression is given by 
\begin{equation}
F_0(E_e,Z)=4(2p_e R_{\rm box})^{2\gamma-2}\frac{\abs{\Gamma(\gamma+i\eta)}^2}{\abs{\Gamma(2\gamma +1)}^2}\text{e}^{\pi\eta}\,,
    \label{eq:F_0}
\end{equation}
with $\gamma\equiv\sqrt{1-(\alpha Z)^2}$ and $\eta\equiv(\alpha Z E_e)/p_e$, where $\alpha$ is the fine-structure constant. The additional term $L_0(E_e,Z)$ corrects for the finite size of the nucleus considering a box density with equivalent radius $R_{\rm box}=\sqrt{5/3}R_{\rm ch}$, while $U(E_e,Z)$ is introduced to improve the nuclear distribution from a box density to a more realistic Fermi distribution. Finally, $S(E_e,Z)$ introduces a correction to the outgoing wave function due to atomic screening~\cite{Hayen:2017pwg, SalvatPhysRevA.36.467, radial}. We have checked that by considering a nucleus with a uniform density and evaluating the Fermi function at the origin, we retrieve the tabulated values of the Fermi function from Behrens \textit{et al.}~\cite{Behrens}, expressed as the product of the first two terms of Eq.~(\ref{eq:Fermi function_lit}), namely $F_0$ and $L_0$. 

\begin{figure}[t]
    \centering
\includegraphics[width=1\linewidth]{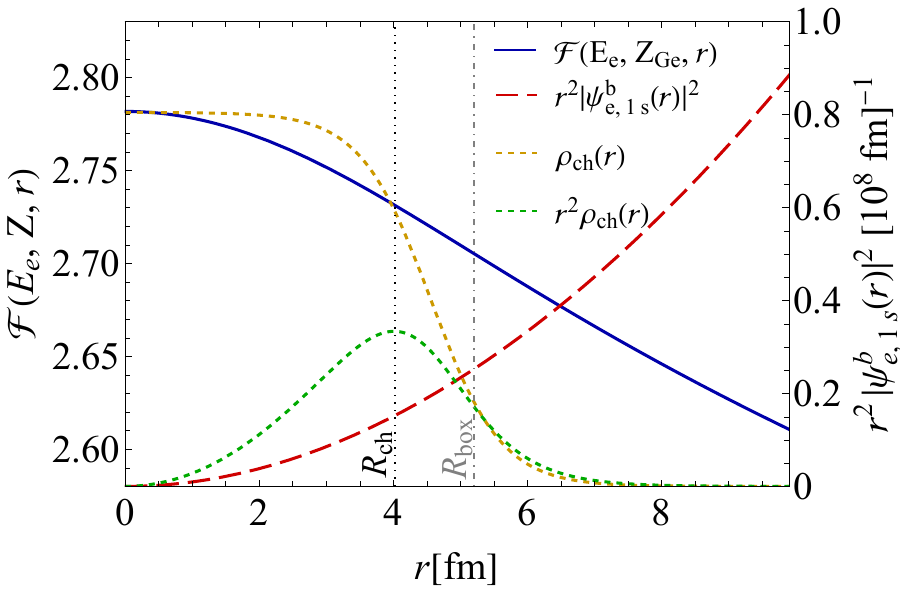}
    \caption{Fermi function for $^{71}$Ge as a function of the radial distance (blue solid line), as derived in this work in Eq.~(\ref{eq:fermi}) using the exact DHFS $g_\kappa(r)$ and $f_\kappa(r)$ wave functions for $E_e=1.031\,\mathrm{MeV}$. We also show the charge density of $^{71}$Ge, $\rho_{\rm ch}(r)$ (gold dashed line) in arbitrary units (also shown in green multiplied by $r^2$) and the electron density for a bound electron in the 1$s$ state of $^{71}\text{Ge}$ (red long dashed line), as obtained from our numerical code for $C_{\rm ex}=1.25$. The vertical dotted and dash-dotted lines represent $R_{\mathrm{ch}}$ and \mbox{$R_{\mathrm{box}}=\sqrt{5/3}\, R_{\rm ch}$}, respectively.}
    \label{fig:Fermifunction}
\end{figure}

In Fig.~\ref{fig:Fermifunction}, we show our result for the Fermi function calculated using the exact DHFS $g_\kappa(r)$ and $f_\kappa(r)$ wave functions in Eq.~(\ref{eq:fermi}) for $E_e=1.031\,\mathrm{MeV}$. 
Our result has been obtained by calculating all of the contributions to the Fermi function within a unique and coherent framework and is in agreement within 0.2--0.3\% with the values reported in Ref.~\cite{Elliott:2023xkb}, as shown in Table~\ref{tab:Fermi_function} for $r_0=0$. Moreover, we have checked that the choice of $C_{\rm ex}$ does not affect the free electron wave functions, with an impact on the evaluation of the Fermi function below 0.004\%.
In Fig.~\ref{fig:Fermifunction}, we also show the electron density for the $1s$ bound state of $^{71}$Ge, as obtained by our numerical calculation for $C_{\rm ex}=1.25$ and the 2pF distribution used to describe the nuclear density.\\

As a further improvement, we move beyond the conventional approximation of evaluating the electron wave function at a single point, as done in Eqs.~(\ref{eq:Hmibd}) and (\ref{eq:lambdaEC}). 
We propose the adoption of averaged wave functions over the $^{71}{\textrm{Ge}}$ nuclear charge density, in both the calculations of IBD and EC amplitudes.\footnote{The average should be evaluated over the Gamow-Teller transition density, defined as the overlap between the $^{71}$Ge and $^{71}$Ga nuclear wave functions with the interaction Hamiltonian. However, this quantity is not currently accessible from first principles in nuclear theory, and we leave its inclusion for an upcoming work.
}
This modification also results in an averaging procedure of the Fermi function in Eq.~(\ref{eq:fermi}), $\overline{\mathcal{F}}(E^j_e,Z_\text{Ge})$, consistent with the prescription originally suggested by Bahcall~\cite{Bahcall:book} for a pointlike nucleus.\footnote{Bahcall suggested to apply the averaging procedure to $F_0$ considering that the ``electron-capture can occur anywhere within the nuclear volume''~\cite{Bahcall:book}. This is not completely justified, since in the evaluation of $F_0$ the nucleus is assumed to be pointlike and considering that the Fermi function corrects the wave function of the outgoing electron and it does not refer to the electron captured in the EC process. On the contrary, the introduction of exact DHFS electron wave functions, as done in this work, naturally leads to the evaluation of the average of the latter within the nucleus when factoring out the electron wave function from the matrix element as in Eq.~(\ref{eq:Hmibd}).} 
The averaging is done through
\begin{eqnarray}
        &&\overline{\mathcal{F}}(E_e,Z)  = \frac{\int_{}^{}{\rho_{\rm ch}(r)\,\mathcal{F}}(E_e,Z,r)\,\mathrm{d}^3r}{\int_{}^{}{\rho_{\rm ch}(r)\,\mathrm{d}^3r}}\, , 
    \label{eq:fermi_mean}
\end{eqnarray}
where $\rho_{\rm ch}(r)$ is the considered nuclear density distribution.
The effect of the averaging procedure of the Fermi function is also shown in Table~\ref{tab:Fermi_function}.

Using this procedure, the ground-state cross section 
becomes
\begin{equation}
\overline{\sigma}^{\rm db}_{\mathrm{gs}}\!=\!\!\frac{2\pi^2 \ln 2}{\overline{f}_\text{EC}t_{1/2}}\!\left(\frac{2J_\text{Ge}\!+\!1}{2J_\text{Ga}\!+\!1} \right)\!\!
\sum_{j} p^j_e E^j_e\,\overline{\mathcal{F}}(E^j_e,Z_\text{Ge})\mathcal{B}(E^j_e),
\label{eq:sigmaIBDFinale}
\end{equation}
where $\overline{f}_{\rm EC}$ is obtained by replacing the punctual wave function in Eq.~(\ref{eq:fEC}) with its average over the nuclear volume,
calculated as~\cite{SensPhysRev.113.679}
\begin{eqnarray}
        &&\overline{|\psi^{\rm b}_{e,1s}|^2} = \frac{\int_{}^{}{\rho_{\rm ch}(r)\,|\psi^{\rm b}_{e,1s}(r)|^2\,\mathrm{d}^3r }}{\int_{}^{}{\rho_{\rm ch}(r)\,\mathrm{d}^3 r}}\, .
    \label{eq:psi_mean}
\end{eqnarray}
The effect of averaging the electron density is shown in the last row of Table~\ref{tab:wavefunction_comparison}, where we compare the values obtained at a fixed radius, namely, $r_0=0$ and $r_0=\sqrt{5/3}R_{\rm ch}$, with that retrieved after the averaging procedure.

\begin{table}[t]
{\renewcommand{\arraystretch}{1.2} 
\begin{tabular}{l|c|c|c|c}
Source &$E_e\,[\text{MeV}]$ & $F(E_e,Z_{\text{Ge}})$ & $\mathcal{F}(E_e,Z_\text{Ge})$ & $\overline{\mathcal{F}}(E_e,Z_\text{Ge})$\\
\hline
$^{51}$Cr & 1.031 & 2.774 & 2.782 & 2.732 \\ 
& 1.030 & 2.775 & 2.782 & 2.733\\
& 1.025 & 2.779 & 2.786 & 2.736\\
& 0.711 & 3.306 & 3.314 & 3.257\\
& 0.710 & 3.309 & 3.317 & 3.260\\
& 0.705 & 3.330 & 3.338 & 3.281\\
\hline
$^{37}$Ar & 1.092 & 2.734 & 2.741 & 2.692\\ 
& 1.092 & 2.734 & 2.742 & 2.692\\
& 1.089 & 2.736 & 2.743 & 2.694\\
\end{tabular}
}
\caption{Comparison between the Fermi function defined according to
Eq.~(\ref{eq:Fermi function_lit})~\cite{Elliott:2023xkb}, $F(E_e,Z_{\text{Ge}})$, and those calculated with our software according to Eqs.~(\ref{eq:fermi}) and~(\ref{eq:fermi_mean}), $\mathcal{F}(E_e,Z_\text{Ge})$ and $\overline{\mathcal{F}}(E_e,Z_\text{Ge})$, for electron energies $E_e$ relevant to $^{51}\text{Cr}$ and $^{37}\text{Ar}$ neutrino sources. In the second and third columns we use $r_0=0$.}\label{tab:Fermi_function}
\end{table}

When exploiting the detailed balance principle, a key piece of information used is the well-known half-life of the EC decay of $^{71}\textrm{Ge}$. 
In this work, four independent measurements of the half-life are employed, namely,
\begin{equation}
\begin{split}
t_{1/2}\left[^{71}\mathrm{Ge}\right]&=\left\{ \begin{array}{cr} 11.43\pm 0.03~\mathrm{d} & \text{\cite{PhysRevC.31.666}}, \\
11.46\pm 0.04~\mathrm{d} & \text{\cite{Collar:2023yew}}, \\ 11.468\pm 0.008~\mathrm{d} &\text{ \cite{PhysRevC.109.055501}}, \\
11.4645\pm 0.0036~\mathrm{d} & \text{\cite{newGelifetime}}; \end{array}\right.
\end{split}
\label{eq:half}
\end{equation}
of which a weighted average has been taken, resulting in $t_{1/2} = 11.465 \pm 0.003~\text{d}$.\\
Also needed is the $Q$-value of the EC process, which is given by the $^{71}\text{Ge}- ^{71}\!\text{Ga}$ atomic mass difference 
\begin{equation}
 Q_\mathrm{EC}=\! M[^{71}\mathrm{Ge}]\!-\!M[^{71}\mathrm{Ga}]\!=\! 232.47 \pm 0.09 \mathrm{~keV},  \label{Qvalue}
\end{equation}
obtained from an average of five different experimental values~\cite{Qvalue}.
The neutrino energy in the EC process depends on the binding energies of the involved $^{71}$Ga states. For a K-shell electron capture, using \mbox{$E_\text{bind}^{\rm K} = 10.367\,\text{keV}$}~\cite{BindingEnergy}, the corresponding neutrino energy is \mbox{$E_{\nu}^{\text{1s}}= 222.10\pm0.09\,\text{keV}$}.
Finally, another important input is the experimentally measured
electron-capture probabilities for the L, M, and K
shells. We combine the measurements of $P_\mathrm{L}/P_\mathrm{K}$ and $P_\mathrm{M}/P_\mathrm{K}$ ratios from Refs.~\cite{Collar:2023yew, PhysRevLett.125.141301} along with the measurements provided in Table~XIV of Ref.~\cite{Bambynek}, obtaining 
\begin{equation}
   \frac{P_\text{L}}{P_\text{K}}\!= 0.1174 \pm 0.0009\, , 
    \frac{P_\text{M}}{P_\text{K}}\!= 0.0188 \pm 0.0004 ,
    \label{eq:PL/PK}
\end{equation}
with associated atomic binding energies 
\mbox{$E_\text{bind}^{\rm L}= 1.2990\,\mathrm{keV}$} and $E_\text{bind}^{\rm M}= 0.1595\,\mathrm{keV}$ for the 
L and M
shells of $^{71}\text{Ga}$~\cite{BindingEnergy}, respectively.

To calculate $f_\text{EC}$, following Eq.~(\ref{eq:fEC}), we take into account the so-called overlap and exchange correction, $\epsilon_o^{1s}$, given that we write the total rate
in terms of the 1s capture rate. We compute the average of the values reported in Refs.~\cite{Bahcall63,VATAI1970}, as previously done in Ref.~\cite{Elliott:2023xkb}, yielding to $\epsilon_o^{1s} = -0.013 \pm 0.007$. We also consider an additional term due to the weak magnetism correction, $[1+\epsilon_\text{q}]_\text{EC}$, with 
$\epsilon_\text{q} = ( 4.9 \pm 0.7)\,\times 10^{-4}$~\cite{Elliott:2023xkb}.\\

In particular, using all these inputs in Eq.\,\eqref{eq:fEC} and after calculating the average of the electron density at the nucleus as in Eq.~(\ref{eq:psi_mean}), we obtain
\begin{equation}
    \overline{f}_\text{EC} = 0.0227 \pm 0.0002 \,,
    \label{eq:fEC number}
\end{equation}
in units of $(\hbar/m_e c)^3$. The latter $\overline{f}_\text{EC}$ is used to extract the nuclear matrix element from the experimental measurement of $t_{1/2}$, as in Eq.~(\ref{eq:matrice EC}). Usually, the numerical value is expressed in terms of the Gamow-Teller transition strength for the IBD process, which results in
\begin{eqnarray}
    B\mathrm{_{GT}}(\mathrm{gs}) = \frac{|\mathcal{M}_{\rm nuc}^{\rm {IBD}}|^2}{2J_{\rm Ga}+1}= 
    0.0859 \pm 0.0006\,,
    \label{eq:BGT number}
\end{eqnarray}
when the detailed balance is applied.

In Table~\ref{tab:cross_section_mean}, we report the ground-state cross section for the IBD process for $^{51}$Cr and $^{37}$Ar neutrino sources. 
The average procedure reduces the ground-state cross section by 2.8\%, indicating a non-negligible correction that should be taken into account.
Moreover, the reported uncertainty on the cross section includes also a systematic uncertainty due to the choice of the exchange potential parameter, estimated as the maximum difference between the cross section calculated for $C_{\rm ex}=1$ and $C_{\rm ex}=3/2$ with respect to that obtained for our benchmark value of $C_{\rm ex}=1.25$. This contribution corresponds to about 0.4\%.
\begin{table}[t]
\resizebox{0.95\columnwidth}{!}
{\renewcommand{\arraystretch}{1.2} 
\begin{tabular}{l|c|c|c|c}
     Source &  
    $\sigma^{\text{db}}_\text{gs} $ &
   $\overline{\sigma}^{\text{db}}_\text{gs} $ & $\overline{\sigma}^{\text{db}}_\text{(p,n)}$ &
   $\overline{\sigma}^{\text{db}}_{(^{3}\text{He},^{3}\text{H})}$ \\
     \hline
 $^{51}\textrm{Cr}$  & $5.42\pm 0.06$ & $5.28\pm 0.06$ & $5.57^{+0.28}_{-0.07}$ & $5.73 \pm 0.17$ \\
  $^{37}\textrm{Ar}$ & $6.50\pm 0.07$ & $6.32\pm 0.07$ & $6.71^{+0.35}_{-0.09}$ & $6.93 \pm 0.22$ \\
\end{tabular} 
}
\caption{Ground-state cross sections for neutrino capture on $^{71}$Ga for $^{51}$Cr and $^{37}$Ar neutrino sources compared to those obtained with two different choices for excited-state contributions, in units of $\left[ 10^{-45}\,\text{cm}^2\right]$. The overline refers to the average over the nuclear volume of the bound-state wave function and the Fermi function.}
\label{tab:cross_section_mean}
\end{table}

To obtain the total cross section, it is necessary to account for the contribution of the excited states corresponding to the transitions $3/2^-\rightarrow5/2^-$ at 175 keV and $3/2^-\rightarrow3/2^-$ at 500 keV~\cite{Elliott:2023xkb},
\begin{equation}
\sigma\!=\!\sigma_\text{gs}\!\bigg[\! 1 + \xi\!\left(5/2^-\!\right)\!\frac{B_{\text{GT}}\left(5/2^-\!\right)}{B_{\text{GT}}(\text{gs})}
+\xi\!\left(3/2^-\!\right)\!\frac{B_{\text{GT}}\left(3/2^-\!\right)}{B_{\text{GT}}(\text{gs})}\! \bigg],
    \label{eq:total cross section}
\end{equation}
where $\xi\left(5/2^-\right)$ and $\xi\left(3/2^-\right)$ are the phase space coefficients, which take the values $\xi\left(5/2^-\right) = 0.669$ and $\xi\left(3/2^-\right) = 0.220$ for the $^{51}\textrm{Cr}$ neutrino source and $\xi\left(5/2^-\right) = 0.696$ and $\xi\left(3/2^-\right) = 0.264$ for the $^{37}\textrm{Ar}$ one. These values are obtained as the ratio of the excited-state cross section to the ground-state one, if $B_{\rm GT}$ is assumed to be the same, as done in Refs.~\cite{Elliott:2023xkb, Bahcall:1997eg}.  
The Gamow-Teller strength values for the excited contributions were obtained, following Ref.~\cite{Elliott:2023xkb}, from the experimental measurement of forward-angle (p,n) scattering~\cite{KrofcheckPhysRevLett.55.1051} and from ($^{3}\text{He},^{3}\text{H}$) scattering~\cite{FREKERS2011134}, leading to different estimations of about 6\,\% and 9\,\%, respectively. In both cases, we report the results in Table~\ref{tab:cross_section_mean} for $^{51}$Cr and $^{37}$Ar neutrino sources. 
Such values can be compared to those reported by the most recent calculation available in the literature~\cite{Elliott:2023xkb}. The ground-state cross sections agree within 0.8\%, while, after applying the averaging procedure, our total cross sections result to be approximately 2\% smaller than those reported in Ref.~\cite{Elliott:2023xkb}.

\section{Reevaluation of the anomaly}
Our calculation of the IBD cross section can be used to determine the number of expected IBD events, $N_\mathrm{cal}$, by considering the neutrino source activity, the atomic density of the target, the average neutrino path length through the absorption material, and the exposure~\cite{PhysRevC.59.2246}. This value can be compared with the experimental measurements from GALLEX~\cite{ANSELMANN1992376,1999127,Kaether:2010ag,GNO:2005bds}, SAGE~\cite{SAGE:2009eeu,doi:10.1142/9789811204296_0002}, and BEST~\cite{Barinov:2021asz,PhysRevC.105.065502}, $N_\mathrm{exp}$, leading to a ratio $R=N_\mathrm{exp}/N_\mathrm{cal}$. Alternatively, one can directly retrieve the ratios $R$ for each experiment, which have been determined using a reference cross section~\cite{Bahcall:1997eg}, by rescaling them to correct for the difference between our cross section and the reference one. The two procedures lead to the same results, which are reported in Table~\ref{tab:Rvalues}. 
We define the following $\chi^2$ function:
\begin{equation}
\chi^2\!=\!\min_\eta\!\Bigg[\!\sum_{k,j}(R_k-\eta \overline{R})(V^{-1})_{kj}(R_j-\eta \overline{R})+\!\left(\frac{1\!-\!\eta}{\Delta\eta}\right)^{2}\Bigg],
    \label{eq:chisquareR}
\end{equation}
where the sum runs over all the experimental measurements $k,j$ and the nuisance parameter $\eta$ is constrained, in the case of the (p,n) excited contribution, by 
\begin{equation}
\begin{split}
\Delta\eta&=\left\{ \begin{array}{cr} \Delta\eta_+(\mathrm{p,n})=0.050 &  \text{for}\,\eta>1, \\
\Delta\eta_-(\mathrm{p,n})=0.013 & \text{for}\,\eta<1, \end{array}\right.
\end{split}
\label{eq:nuis}
\end{equation}
and $\Delta\eta(^3\mathrm{He},^3\mathrm{H})=0.031$ for the ($^{3}\text{He},^{3}\text{H}$) one.

\begin{table}[h]
\resizebox{0.95\columnwidth}{!}
{\renewcommand{\arraystretch}{1.2} 
\begin{tabular}{c|c|c|c}
    & $R_{\rm gs}$& $R_{\rm (p,n)}$& $R_{\rm (^{3}\text{He},^{3}\text{H})}$\\\hline
    \multirow{2}{4em}{GALLEX Cr1} & $1.049\pm0.121$ & $0.994\pm0.115$ & $0.966\pm0.112$\\\
    & & &\\\hline
    \multirow{2}{4em}{GALLEX Cr2} & $0.894\pm0.121$ &$0.847\pm0.115$ & $0.823\pm0.112$\\ & & &\\\hline
    SAGE Cr & $1.045\pm0.132$ & $0.991\pm0.125$ & $0.963\pm0.122$\\\hline
    SAGE Ar & $0.875\pm0.105$ & $0.824\pm0.099$ & $0.798\pm0.096$\\\hline
    \multirow{3}{4em}{BEST Cr-R1} & \multirow{3}{5 em}{$0.870\pm0.032(\text{unc})\pm0.023(\text{cor})$} & \multirow{3}{5 em}{$0.825\pm0.031(\text{unc})\pm0.022(\text{cor})$} & \multirow{3}{5 em}{$0.802\pm0.030(\text{unc})\pm0.021(\text{cor})$}\\ & & &\\ & & &\\\hline
    \multirow{3}{4em}{BEST Cr-R2} & \multirow{2}{5 em}{$0.843\pm0.031(\text{unc})\pm0.027(\text{cor})$} & \multirow{3}{5 em}{$0.799\pm0.029(\text{unc})\pm0.025(\text{cor})$} & \multirow{3}{5 em}{$0.777\pm0.029(\text{unc})\pm0.025(\text{cor})$}\\ & & &\\ & & &\\\hline
\end{tabular} 
}
\caption{Ratio between the experimental number of IBD events and the predicted one for each experimental measurement considered, together with the corresponding uncertainties, divided, when necessary, between correlated (cor) and uncorrelated (unc).}
\label{tab:Rvalues}
\end{table}

\begin{figure}[t]
    \centering
    \includegraphics[width=\linewidth]{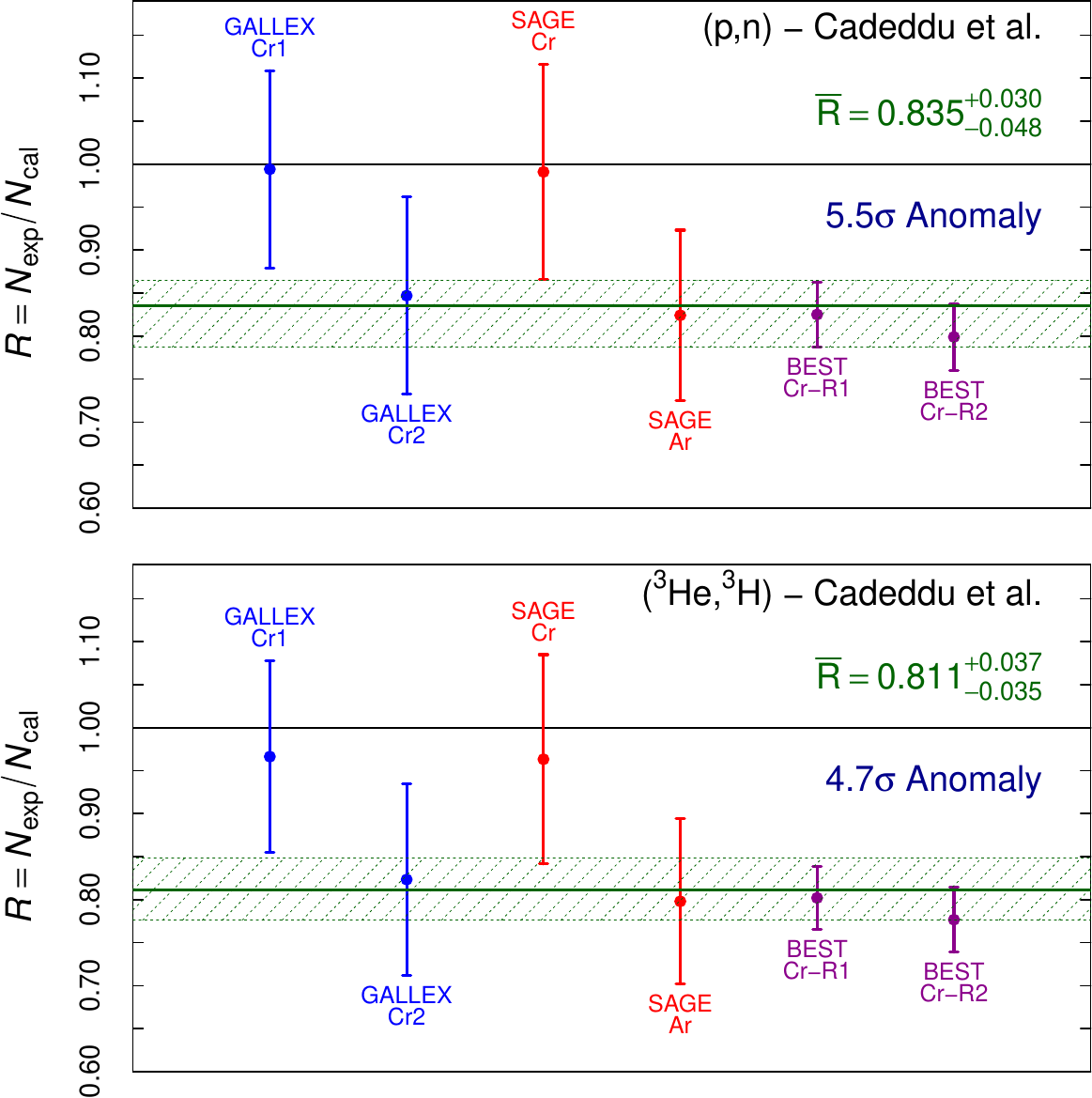}    \caption{Ratios between the experimental values and our theoretical estimations of the event rates in gallium experiments,
when considering the (p,n) (upper panel)
or the ($^{3}\text{He},^{3}\text{H}$) (lower panel) excited-state contribution.
Also shown are the best-fit value of $\overline{R}$ (green line)
and its 1$\sigma$ uncertainty (green band).}
   \label{fig:statusR}
\end{figure}

\begin{figure}[t]
    \centering
    \includegraphics[width=\linewidth]{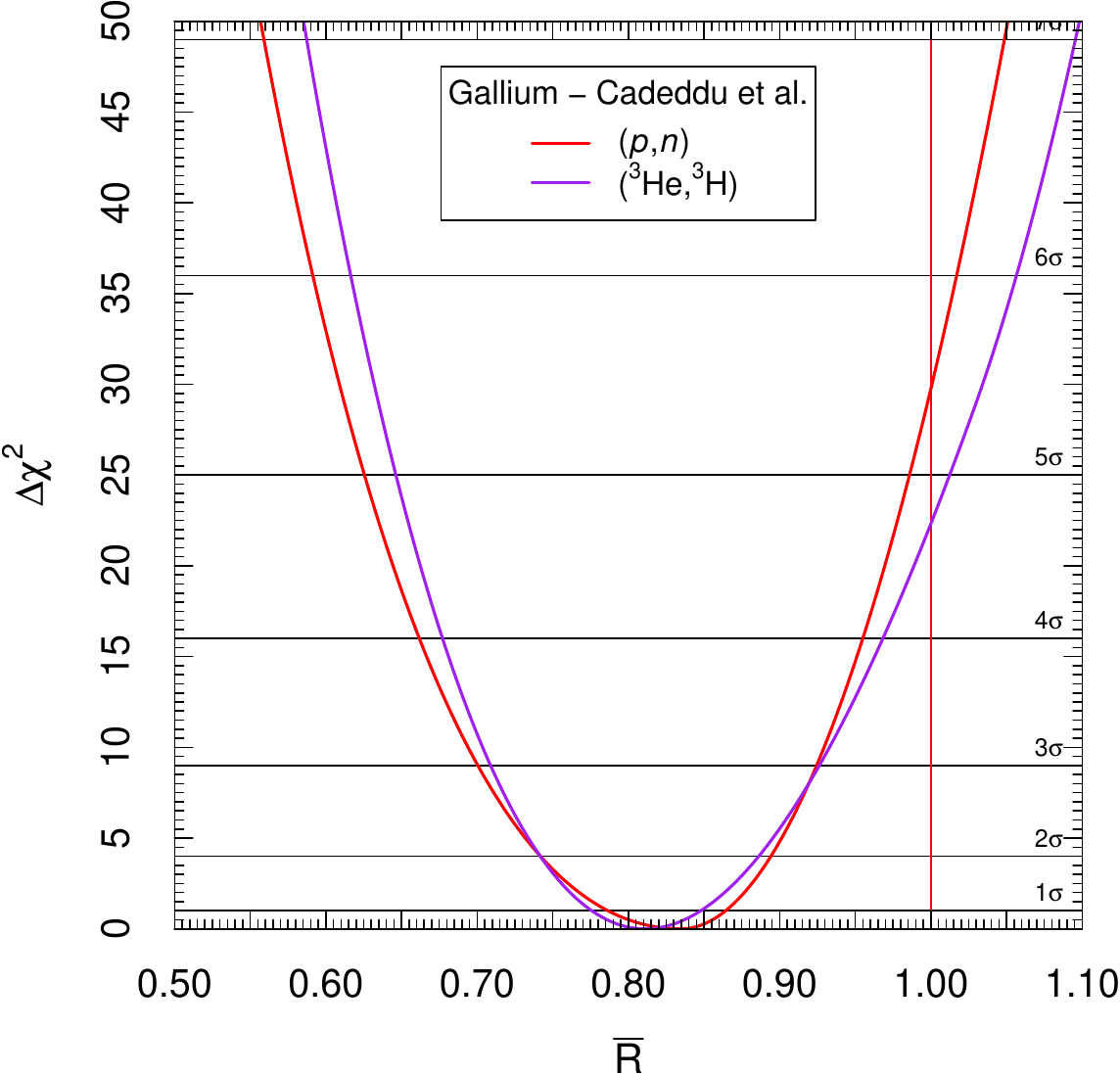}    \caption{$\Delta\chi^2 = \chi^2 - \chi^2_{\text{min}}$
    as a function of $\overline{R}$ obtained from the combined analysis
    of the gallium data when considering the (p,n) (red line)
or the ($^{3}\text{He},^{3}\text{H}$) (purple line) excited-state contribution.}
   \label{fig:rav-chi-plt}
\end{figure}

The covariance matrix $V$ is obtained from the uncertainties on the $R_k$ ratios.
The resulting best-fit value for all the experiments is found to be $\overline{R}=0.835^{+0.030}_{-0.048}$, when considering the (p,n) excited contribution, see Fig.~\ref{fig:statusR} (upper panel). 
Taking into account the nonparabolic behavior of the $\chi^2$,
shown in Fig.~\ref{fig:rav-chi-plt},
we obtain a $5.5\sigma$ anomaly. With the ($^{3}\text{He},^{3}\text{H}$) excited contribution
we obtain $\overline{R}=0.811^{+0.037}_{-0.035}$,
also shown in Fig.~\ref{fig:statusR} (lower panel),
and a smaller $4.7\sigma$ anomaly. These results have been obtained by improving
the calculation method used in Ref.~\cite{Giunti:2022xat}
by removing the model uncertainties from the BEST data, to avoid double counting.
These uncertainties are correlated among the experiments
and are taken into account in the $\chi^2$ with the $\eta$ nuisance parameter.
We also took into account the correlation of the systematic uncertainties of the
BEST data.
Moreover, to provide a conservative estimate of the anomaly, we reevaluate it considering the ground-state cross section in Eq.~(\ref{eq:sigmaIBDFinale}), whose best fit of the data is found to be $\overline{R}=0.881 \pm 0.031$, which corresponds to a $3.8\sigma$ anomaly.

Our result is in good agreement with previous ones~\cite{Elliott:2023xkb,Haxton:2025hye,Bahcall:1997eg}, although with differences in the procedure. In fact, the averaging procedure, which leads to a lower cross section value, compensates for the refined statistical analyses, leading to a similar significance of the anomaly. 
It is worth stressing that in this work we use the same software and underlying assumptions when evaluating the electron wave functions in both the Fermi function and $\overline{f}_\text{EC}$ calculations in Eq.~(\ref{eq:sigmaIBDFinale}). Entering as a ratio, any systematic shift should cancel in the cross section determination, making the result more robust and self-consistent. 

\begin{figure}[h]
    \centering
    \includegraphics[width=\linewidth]{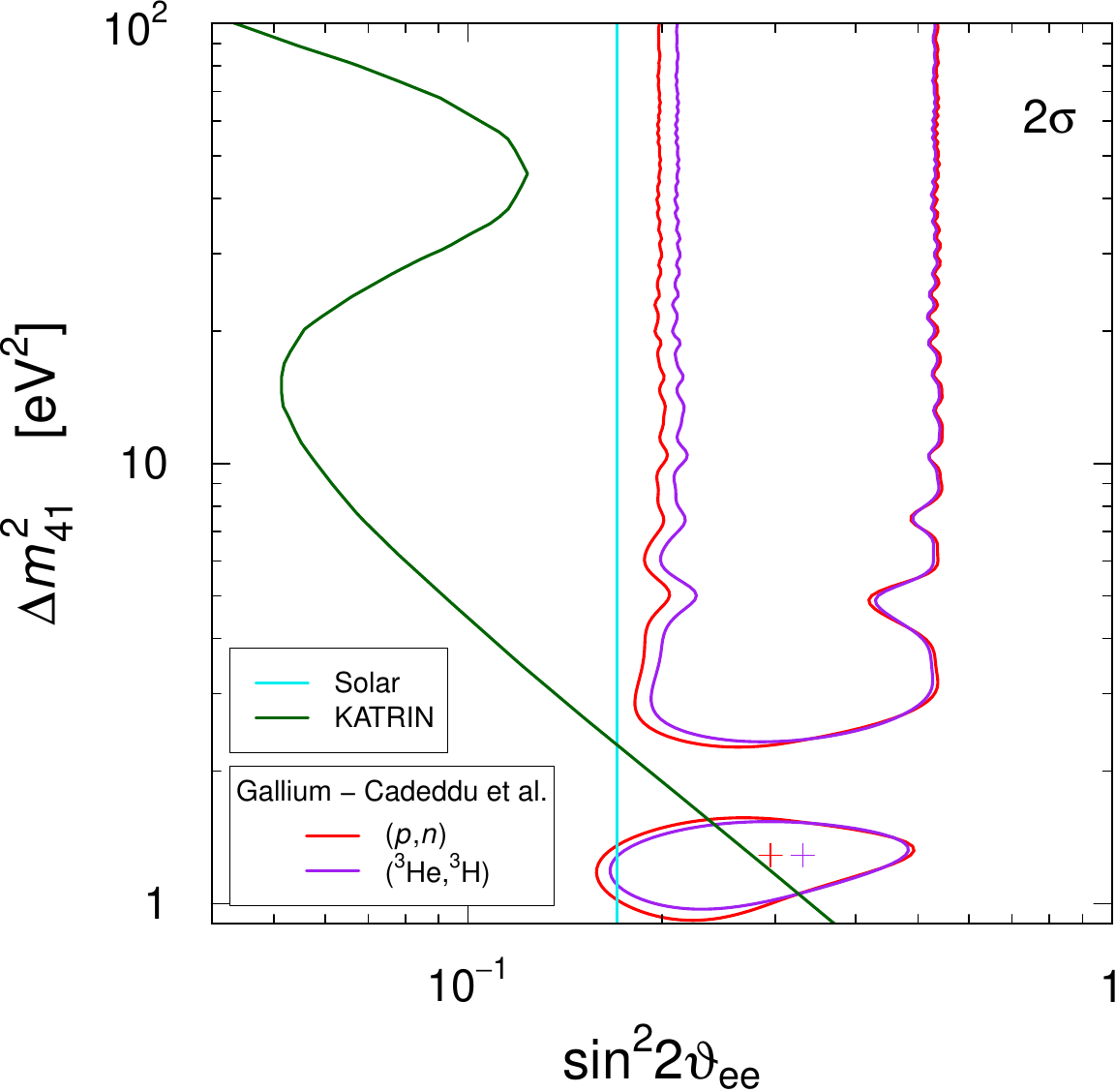}
    \caption{Contours of the $2\sigma$ allowed regions in the
    plane of the effective oscillation parameters
    of 3+1 active-sterile neutrino mixing
    obtained from the combined analysis of the gallium data when considering the (p,n) (red line) or the ($^{3}\text{He},^{3}\text{H}$) (purple line) excited-state contribution.
    The crosses indicate the best-fit points.
    Also shown are the $2\sigma$
    solar bound $\sin^2 2\vartheta_{ee} \lesssim 0.17$~\cite{Gonzalez-Garcia:2024hmf} (cyan line) and the 95\% C.L. bound of the KATRIN experiment~\cite{KATRIN:2025lph} (green line).
    }
    \label{fig:sterile}
\end{figure}

\section{Sterile Neutrinos}
The simplest and most considered explanation of the gallium anomaly is the hypothesis of
short-baseline active-sterile neutrino oscillations
(see the recent review in Ref.~\cite{Elliott:2023cvh}).
Figure \ref{fig:sterile} shows the allowed regions in the space of the effective parameters
of 3+1 short-baseline active-sterile oscillations
(see Ref.~\cite{Giunti:2019aiy}).
Our results confirm those obtained in previous works
\cite{Barinov:2021asz,Barinov:2022wfh,Barinov:2021mjj,Berryman:2021yan,Giunti:2022xat,Giunti:2022btk,Elliott:2023xkb,Giunti:2023kyo},
which show that the gallium data,
including the recent and most precise BEST data,
can be explained by
short-baseline active-sterile neutrino oscillations
with a large mixing
($0.16 \lesssim \sin^2 2\vartheta_{ee} \lesssim 0.54$ at $2\sigma$,
for $\Delta{m}^2_{41} \gtrsim 0.98 \, \text{eV}^2$).
Such large mixing is in tension with the bounds obtained from the analysis
of reactor antineutrino data
\cite{Giunti:2021kab,Giunti:2022btk,STEREO:2022nzk,Machikhiliyan:2022muh,PROSPECT:2024gps},
the solar neutrino bound
\cite{Goldhagen:2021kxe,Gonzalez-Garcia:2024hmf},
and the KATRIN neutrino mass measurement
\cite{KATRIN:2025lph}.
The solar and KATRIN bounds are shown in Fig.~\ref{fig:sterile} for comparison.
The reactor bound is more complicated,
because it depends on the reactor antineutrino flux model
and on the choice of the dataset
\cite{Giunti:2021kab,Giunti:2022btk}.
These tensions disfavor the sterile neutrino explanation of the
gallium anomaly,
which is, however, still considered
because so far nobody has found a plausible alternative explanation.\footnote{Exotic explanations have been proposed in
Refs.~\cite{Arguelles:2022bvt,Hardin:2022muu,Banks:2023qgd,Farzan:2023fqa,Brdar:2023cms}.
Some have been shown to be in tension with the reactor rate data
\cite{Giunti:2023kyo}.}

\section{Conclusions}
In this work, we critically reassessed the status of the gallium anomaly. By employing exact Dirac-Hartree-Fock-Slater wave functions for both the bound and continuum electron states, a new averaging procedure, and a refined statistical analysis, we reevaluated the neutrino capture cross section on $^{71}$Ga, showing that a $5.5\sigma$ discrepancy is now achieved with the (p,n) excited-state contribution,
and $4.7\sigma$ with the ($^{3}\text{He},^{3}\text{H}$) excited-state contribution. Such a result was obtained by using the same software and underlying assumptions when evaluating the electron wave functions in both the Fermi function and electron density at the nucleus for the EC process, which enter as a ratio in the cross section. Thus, we avoided possible systematic shifts, making the result more robust and self-consistent.
We also presented the updated allowed region in the space of the
effective parameters of 3+1 active-sterile neutrino mixing.
We confirmed the need for the explanation of the gallium data
of a large active-sterile mixing,
which is in tension with the results of reactor antineutrino experiments
\cite{Giunti:2021kab,Giunti:2022btk,STEREO:2022nzk,Machikhiliyan:2022muh,PROSPECT:2024gps},
with the solar neutrino bound
\cite{Goldhagen:2021kxe,Gonzalez-Garcia:2024hmf},
and the KATRIN neutrino mass measurement
\cite{KATRIN:2025lph}.
Hence, the solution of the gallium anomaly remains an open problem,
which may be solved by a new source experiment
with a different detector~\cite{Huber:2022osv,Ciuffoli:2025quh,Chauhan:2025vly}.

\begin{acknowledgements}
The authors gratefully acknowledge J. Ekman, J. Biero\'n, and P. J\"onsson for their valuable insights on GRASP and Mehrdad S. Beni for his helpful clarification regarding the electron density at the nucleus.

\end{acknowledgements}

\section*{Data Availability}
The data that support the findings of this article are openly available as ancillary files on arXiv~\cite{data}.

\bibliography{ref}

\end{document}